\documentclass[aps,prd,superscriptaddress,showpacs,twocolumn,nofootinbib]{revtex4-1}
\usepackage{graphicx}
\usepackage[caption=false]{subfig}
\usepackage{epstopdf}
\usepackage{amsmath}
\usepackage{amsfonts} 
\usepackage{amssymb}
\usepackage{latexsym}
\usepackage{hyperref}
\usepackage[english]{babel}
\usepackage[utf8]{inputenc}
\usepackage[dvipsnames]{xcolor}
\usepackage{slashed}
\usepackage{feynmp}
\usepackage{bm}
\usepackage{bbold}
\usepackage{eufrak}
\usepackage{tabu}
\usepackage{tikz}

\usepackage{bigstrut}

\begin{document}

\title{Bounds on the photon mass via the Shapiro effect in the solar system}

\author{P. C. Malta}\email{pedrocmalta@gmail.com}
\affiliation{R. Antonio Vieira 23, 22010-100, Rio de Janeiro, Brazil}

\author{C. A. D. Zarro}\email{carlos.zarro@if.ufrj.br}
\affiliation{Instituto de F\'isica, Universidade Federal do Rio de Janeiro, RJ 21941-972 – Brazil}


\begin{abstract}

We study the effects of a finite mass for the photon on its propagation in a weak gravitational field. In particular, we analyse the gravitational time delay, also known as the Shapiro effect. We work in isotropic coordinates in the weak-field limit and find that the mass-dependent corrections enhance the gravitational time delay. Doppler-tracking data from the Cassini mission allow us to set an upper bound on the photon mass, namely $m_\gamma < 4.9 \times 10^{-7} \, {\rm eV/c^2}$ at $95\%$~CL. We also discuss next-generation solar-system tests of general relativity that could improve this upper limit, potentially by a factor of ten. Though not competitive with the currently best limits, our bound is at the ballpark of earlier ones based on the gravitational bending of light by the Sun.

\end{abstract}
\maketitle


\section{Introduction}  \label{sec_intro}
\indent

The photon is the paradigm for a massless particle. In fact, the invariance of electromagnetism under local gauge transformations requires $m_\gamma = 0$, though the photon may indeed acquire an effective mass in certain circumstances~\cite{Skobelev, Cover, supercond}. The fundamental masslessness of the photon is a theoretically pleasing hypothesis, usually taken for granted, but it must be thoroughly tested. The most stringent upper bound currently accepted is $m_\gamma < 10^{-18} \, {\rm eV/c^2}$ extracted from data from the Voyager missions~\cite{Ryutov}; for reviews, see the latest edition of the Particle Data Group~\cite{PDG}, as well as refs.~\cite{Tu, Nieto, Okun, Goldhaber,Goldhaber2}.

Measuring the photon mass is only possible if the dimensions of the system are very large or the measurement is very precise~\cite{Goldhaber, Goldhaber2, Nieto}, since the photon mass only appears squared and effects are $\mathcal{O}(m_\gamma^2 d^2)$, where $d$ is the typical dimension of the system. This severely limits the range of feasible tests, since experiments involving atomic physics, which reach astounding precision, are bound to atomic dimensions, thus leading to relatively weak upper limits~\cite{Accioly2}. At the intermediate planetary scale the limits profit from larger length scales, but still suffer from relatively low-precision, or noisy data~\cite{Kroll, Malta}. Finally, at the other extreme, solar-system tests benefit from astronomical distances and place the currently strongest upper bounds~\cite{Ryutov, Retino}.

Here we study the effect of the photon mass on the gravitational time delay, first suggested by Shapiro in 1964 as a ``fourth test of general relativity"~\cite{Shapiro1}. His idea was to send signals from radio telescopes on Earth towards Mercury or Venus when these are close to superior conjunction, {\it i.e.}, passing almost behind the Sun. According to Shapiro's calculations, the round-trip time of the signal would be longer than expected in the absence of the Sun: for grazing incidence at the solar limb, the extra travel time due to gravity can reach $\sim 200 \, {\rm \mu s}$ in the inner solar system.

The measurements were conducted using the Arecibo and Haystack telescopes with dedicated radio systems allowing the detection of the faint echoes ($\approx 10^{27}$ weaker than the emitted signals)~\cite{Shapiro2, Shapiro3}. Curvature effects are characterized by Eddington's $\gamma$ parameter, which is unity in standard general relativity~\cite{Poisson, Will_book}, and the data analysis yielded $\gamma = 1.03 \pm 0.04$, thus confirming the predictions of general relativity to within a few percent. Other similar analyses were undertaken using ranging data from the Mariner landers~\cite{Mariner1} and the Viking spacecraft~\cite{Viking, Viking2, Viking3}, both sent to Mars, determining $\gamma$ to within $3\%$ and $0.2\%$ accuracy, respectively. The most precise determination of $\gamma$ so far stems from Doppler-tracking data from the Cassini spacecraft on its way to Saturn, which yielded $|\gamma - 1| = 2.3 \times 10^{-5}$~\cite{Cassini_1, Cassini_2}. These results will be explored in Sec.~\ref{sec_exp_sens} to constrain the photon mass.

Quite generally, the search for deviations of general relativity, or new-physics effects in general, can profit greatly from experiments in space~\cite{Book_space}. Tests of the universality of free fall, for example, may be performed over longer time periods, thus reaching higher precision levels. Similarly, interferometric experiments can benefit from much longer interaction times, as well as longer distances, both helping to enhance possible signals. The absence of seismic vibrations and atmospheric disturbances, which impart non-gravitational accelerations to sensors, are further advantages of space-based experiments. Also for time-delay measurements the space environment is ideal, since precise time keeping is crucial when minute height and time inaccuracies may impact the results. In space these variables can be particularly well controlled, thus allowing for very precise measurements.

So far all measurements of the Shapiro effect were performed using ground-based antennas sending and receiving signals to spacecraft at interplanetary distances. Now, however, the technology necessary to measure time delays to picosecond accuracy in space is mature ({\it e.g.}, with highly stable laser sources, ultra-precise atomic clocks, drag-free control, etc~\cite{Book_space}). In this context, we also analyse a couple of promising space missions which include time-delay measurements as their scientific targets aiming at considerably improving the current limit on $|\gamma - 1|$.

This paper is organized as follows: in Sec.~\ref{sec_grav_effects} we set up the theoretical basis of our analysis and find the gravitational time delay in the case of massive photons. Next, in Sec.~\ref{sec_exp_sens} we use data from the Cassini time-delay experiment to extract an upper bound on the photon mass. Finally, in Sec.~\ref{sec_conclusions} we present our conclusions and briefly comment on future prospects from upcoming space-based missions. We use SI units throughout.


\section{Gravitational effects}  \label{sec_grav_effects}
\indent

We are interested in the behavior of a heavy photon -- a massive and electrically neutral Abelian spin-1 boson -- propagating in a weak gravitational field. The dynamics of this particle is governed by the de Broglie-Proca action coupled to gravity~\cite{dB1, dB2, dB3, Proca1, Proca2}:
\begin{equation} \label{eq_lag_proca}
S = \int d^4 x \sqrt{-g}\left[ -\frac{1}{4\mu_0} F^{\mu\nu} F_{\mu\nu} + \frac{\mu_\gamma^2}{2\mu_0} A_\mu A^\mu \right] \, ,
\end{equation}
where $\mu_\gamma = m_\gamma c/\hbar$ is the reciprocal reduced Compton wavelength and $\mu_0$ is the vacuum permeability. The anti-symmetric field-strength tensor is $F_{\mu\nu} = \partial_\mu A_\nu - \partial_\nu A_\mu$ and $g$ is the determinant of the metric tensor.

Before starting we must decide on the coordinates to describe the spacetime curvature caused by a central mass (here we neglect the its finite size). While standard Schwarzschild coordinates are common, here we employ heliocentric isotropic spherical coordinates with which the interval squared is given by
\begin{equation}\label{eq_ds2}
ds^2 = \left( \frac{1 - \frac{GM}{2c^2 r}}{1 + \frac{GM}{2c^2 r}}  \right)^2 c^2 dt^2 - \left( 1 + \frac{GM}{2c^2 r} \right)^4 \left( dr^2 + r^2 d\Omega^2 \right) 
\end{equation}
with $d\Omega^2 = d\theta^2 + \sin^2\theta d\phi^2$, where $\theta$ and $\phi$ are the polar and azimuthal angular coordinates, respectively. Henceforth we take the Sun to be the central mass\footnote{$M_\odot = 1.99 \times 10^{30}$~kg, $R_\odot = 6.96 \times 10^8$~m.}, so $M = M_\odot$. As usual, the geodesics will be parametrized by the affine parameter $\lambda = \tau$, the proper time of the massive photon. Moreover, the constraint $g_{\mu\nu} \dot{x}^\mu \dot{x}^\nu = c^2$ must be enforced; here the dot indicates differentiation with respect to the affine parameter.

Isotropic coordinates are typically used in astrometric calculations and are particularly suited for comparison with observational data. Also important in our case is the fact that the post-Newtonian terms in the context of a parametrized post-Newtonian (PPN) expansion~\cite{Will_book, Will_exp, Turyshev} are expressed in these coordinates. Our choice therefore simplifies the comparison between our results including a finite photon mass and the PPN expressions containing Eddinton's $\gamma$ parameter, cf. App.~\ref{app_A}.



\subsection{Deflection angle}  \label{sec_angle}
\indent

One of the three classic tests of Einstein's general relativity is the deflection of light by a massive object, which is twice as large as the Newtonian prediction: $\delta_{\rm E} = 2\delta_{\rm N} = 4GM_\odot/c^2 r_0$, where $r_0$ is the distance of closest approach (for the Sun it is the perihelion). Light, typically taken to be exactly massless, is just an extreme test case of the theory, since massive particles are obviously also affected by curvature. We are interested in the latter case, where photons are exceedingly light, but massive nevertheless.

In App.~\ref{app_B} we derived the geodesic equations\footnote{It may be shown that, in the limit of geometric optics~\cite{Harte_2018}, the inclusion of a mass scale does not modify the fact that the constant phase surface is normal to the wave vector, which satisfies a geodesic equation.} and showed that we are allowed to constrain the trajectory to the equatorial plane ($\theta = \pi/2$) without loss of generality. Our target is to obtain an expression for $d\phi/dr$ as a function of the radial coordinate and, to this end, we start with eq.~\eqref{eq_phidot_2} writing it as
\begin{equation} \label{eq_omega}
\dot{\phi} = \frac{K}{r^2 \left( 1 + \frac{\mathcal{M}}{2r} \right)^4} \, ,
\end{equation}
where $\mathcal{M} = GM_\odot/c^2 \approx 1.5$~km.

The constant of integration $K$ may be determined by evaluating the constraint~\eqref{eq_constraint_2} at the perihelion, $r = r_0$, where $\dot{r} = 0$. By doing so, we find $K^2 = c^2 r_0^2 F(\mathcal{M}/2r_0)$ with $F(y)$ defined as
\begin{equation} \label{eq_F}
F(y) = \left( 1 + y \right)^4 \left[ \frac{1}{\alpha^2} \frac{ \left( 1 + y \right)^2 }{ \left( 1 - y \right)^2 }  - 1  \right] \, 
\end{equation}
and $\alpha = m_\gamma c^2/E$, which may be expressed as
\begin{equation} \label{eq_alpha}
\alpha = 2.4 \times 10^{-9} \, \left( \frac{m_\gamma}{10^{-14} \, {\rm eV/c^2}} \right) \left( \frac{1 \, {\rm GHz}}{f} \right) \, ,
\end{equation}
where we used the quantum-mechanical relation $E = hf$. Plugging eq.~\eqref{eq_omega} into the constraint $g_{\mu\nu} \dot{x}^\mu \dot{x}^\nu = c^2$, cf. eq.~\eqref{eq_constraint_2}, we find 
\begin{equation} \label{eq_r_dot}
\dot{r} = c\frac{r_0}{r} \frac{ \sqrt{F(\mathcal{M}/2r_0)} }{\left( 1 + \frac{\mathcal{M}}{2r} \right)^{4}}  \left[ \left( \frac{r}{r_0} \right)^2 \frac{F(\mathcal{M}/2r)}{F(\mathcal{M}/2r_0)} - 1  \right]^{1/2} \, ,
\end{equation} 
which duly satisfies $\dot{r} = 0$ at $r = r_0$. Finally, dividing eq.~\eqref{eq_omega} by eq.~\eqref{eq_r_dot}, we find
\begin{equation}\label{eq_dphi_dr}
\frac{d\phi}{dr} = \frac{1}{r} \left[ \left( \frac{r}{r_0} \right)^2 \frac{F(\mathcal{M}/2r)}{F(\mathcal{M}/2r_0)} - 1  \right]^{-1/2} \, .
\end{equation}

We are interested in determining the total change in angle as the photon passes by the Sun. Noting that the orbit is symmetric around the perihelion, we only need to integrate the radial coordinate from $r_0$ to infinity and multiply by two. The total angle covered is $\Delta\phi = \pi + \delta$, with $\delta$ describing the gravity-induced deviation of the path from the unperturbed straight line. Changing variables to $q = r_0/r$ and letting $x = \mathcal{M}/2r_0$, we get
\begin{equation}
|\Delta\phi | = 2\int_0^1 dq \left[ \frac{F(qx)}{F(x)}  - q^2  \right]^{-1/2} \, .
\end{equation}
Noting that $x \ll 1$, we may expand the integrand and integrate term by term with the result
\begin{equation}
|\Delta\phi | = \pi + 4 \left( \frac{2 - \alpha^2}{1  - \alpha^2} \right) x + \mathcal{O}(x^2) \, .
\end{equation}
Since $\alpha^2 \ll 1$ by assumption, we finally obtain
\begin{equation} \label{eq_delta}
\delta = \frac{4GM_\odot}{c^2 r_0} \left( 1 + \frac{\alpha^2}{2} \right) \, 
\end{equation}
to first post-Newtonian order. Clearly, the standard result is recovered in the massless limit and the photon mass appears squared, as expected~\cite{Nieto}. It is also clear that a finite photon mass acts to increase the deflection angle, which now depends on the energy of the incoming photon, meaning that different frequencies are deflected at different angles, a dispersive effect absent from the usual massless case.

Equation~\eqref{eq_delta} matches the results from Refs.~\cite{Accioly1, Accioly2} calculated via scattering methods, as well as those from a simplified calculation using a straight-line approximation, cf. App.~\ref{app_C}. In Ref.~\cite{Accioly2} the authors use 43-GHz data from the Very-Long Baseline Array (VLBA)~\cite{Fomalont} and derive $m_\gamma \lesssim 3.5 \times 10^{-5} \, {\rm eV/c^2}$, whereas in Ref.~\cite{Accioly1} interferometric techniques are also used, but with 2-GHz radio waves~\cite{Lebach}, to obtain\footnote{In Ref.~\cite{Accioly1} the authors quote $m_\gamma \lesssim 10^{-40} \, {\rm g} \approx 5.6 \times 10^{-8} \, {\rm eV/c^2}$. However, the deviation of the measured angles relative to Einstein's result is $\Delta/\delta_{\rm E} \approx 8 \times 10^{-4}$~\cite{Lebach}, so that using the first equation in their section 3 we get $m_\gamma \lesssim 5.9 \times 10^{-40} \, {\rm g} \approx 3.3 \times 10^{-7} \, {\rm eV/c^2}$, a factor six worse than their stated limit.} $m_\gamma \lesssim 3.3 \times 10^{-7} \, {\rm eV/c^2}$. 




\subsection{Gravitational time delay}  \label{sec_time_delay}
\indent

One of the most interesting theoretical consequences of general relativity is that light signals propagating near a massive object will take a longer time to transit between source and receiver than what would be expected from Newtonian theory alone. This gravitational time delay was first calculated by Shapiro in 1964 in the context of radio signals sent from Earth towards Mercury or Venus close to superior conjunction~\cite{Shapiro1, Shapiro2, Shapiro3}. In this section we calculate the Shapiro effect assuming that the signal is composed of massive photons following a similar strategy as above in the study of the deflection angle.

We want to determine the total (coordinate) time elapsed from emission of the signal on Earth until its arrival at the receiver, so we must find $dt/dr$ as a function of known parameters, in particular the heliocentric radial distances of the objects involved. Once again we constrain the movement to the equatorial plane. We do not need to go back to the geodesic equations: it suffices to divide the constraint equation~\eqref{eq_constraint_2} by $\dot{r}$ to get
\begin{eqnarray} \label{eq_dtdr_1}
\left( \frac{dt}{dr} \right)^2 & = & \frac{\left( 1 + \frac{\mathcal{M}}{2r} \right)^6}{c^2 \left( 1 - \frac{\mathcal{M}}{2r} \right)^2} \left[  1 + r^2 \left( \frac{d\phi}{dr} \right)^2 \right] \nonumber \\
& + & \frac{\left( 1 + \frac{\mathcal{M}}{2r} \right)^2}{\left( 1 - \frac{\mathcal{M}}{2r} \right)^2} \frac{1}{\dot{r}^2}  \, .
\end{eqnarray}

As in Sec.~\ref{sec_angle}, it is useful to change variables to $q = r_0/r$, where $r_0$ is the shortest distance to the Sun, and define $x = \mathcal{M}/2r_0$, so that $\mathcal{M}/2r = qx$. With these changes and plugging in eqs.~\eqref{eq_r_dot} and~\eqref{eq_dphi_dr} for $\dot{r}$ and $d\phi/dr$, respectively, eq.~\eqref{eq_dtdr_1} becomes
\begin{eqnarray}
\frac{dt}{dq} & = & \left( \frac{r_0}{c} \right) \frac{\left( 1 + qx \right)^3}{ q^2 \left( 1 - qx \right)} \frac{1}{\sqrt{H(q,x)}}  \nonumber \\
& \times & \left[ 1 + H(q,x) + \frac{\left( 1 + qx \right)^4}{q^2 F(x) } \right]^{1/2}
\end{eqnarray}
with $F(x)$ given by eq.~\eqref{eq_F} and $H(q,x)$ defined as
\begin{equation} \label{eq_H}
H(q,x) = \frac{1}{q^2} \frac{F(qx)}{F(x)} - 1 \, .
\end{equation}

We must now compute the integral from the emitter or receiver of the signal at $q_i = r_0/r_i$ ($i = e,r$) to the perihelion at $q_p = 1$. Before integrating, it is convenient to first expand the integrand in powers of $\alpha^2 \ll 1$ and in $x  = \mathcal{M}/2 r_0 \ll 1$. Keeping only the leading-order contributions, we get
\begin{eqnarray} 
\Delta t_{\rm i \rightarrow p} & = & \frac{1}{c} \sqrt{r_i^2 - r_0^2} \left( 1 + \frac{\alpha^2}{2} \right) \nonumber \\
& + & \frac{2GM_\odot}{c^3} \sqrt{ \frac{r_i - r_0}{r_i + r_0} } \left( 1 + \alpha^2 \right) \nonumber \\
& + & \frac{2GM_\odot}{c^3} \log\left( \frac{r_i + \sqrt{r_i^2 - r_0^2}}{r_0} \right) \, . \label{eq_Delta_ip}
\end{eqnarray}
The one-way trip is composed of the path from emitter to receiver passing by the perihelion: $\Delta t_{\rm e \rightarrow p} + \Delta t_{\rm r \rightarrow p} $. Due to symmetry, the total coordinate time of the round trip is twice that, therefore

\begin{eqnarray} \label{eq_Delta_t_final}
\Delta t & = & \frac{2}{c} \left[ \sqrt{r_e^2 - r_0^2} + \sqrt{r_r^2 - r_0^2} \right] \left( 1 + \frac{\alpha^2}{2} \right) \nonumber \\
& + & \frac{4GM_\odot}{c^3} \left[  \sqrt{ \frac{r_e - r_0}{r_e + r_0} } + \sqrt{ \frac{r_r - r_0}{r_r + r_0} } \right] \left( 1 + \alpha^2 \right)  \\
& + & \frac{4GM_\odot}{c^3} \log\left[ \frac{ \left( r_e + \sqrt{r_e^2 - r_0^2} \right) \left( r_r + \sqrt{r_r^2 - r_0^2} \right)  }{r_0^2} \right] \, .\nonumber
\end{eqnarray}

The result above is not, however, what a clock on Earth would measure, since eq.~\eqref{eq_Delta_t_final} refers to the coordinate time as measured by a far-away observer. On Earth we must account for the gravitational influence of the Sun, as well as Earth's translational motion. From the line element, cf. eq.~\eqref{eq_ds2}, with $ds^2 = c^2 d\tau^2$ evaluated at $r = r_e$ (with $\dot{r} = 0$) we have
\begin{equation}
\left( \frac{\Delta\tau}{\Delta t} \right)^2 \approx \left( 1 - \frac{2GM_\odot}{c^2 r_e} \right) - \left( 1 + \frac{2GM_\odot}{c^2 r_e} \right) r_e^2 \frac{\dot{\phi}^2}{c^2} \, .
\end{equation}
Here $GM_\odot/c^2 r_e \ll 1$ was used to expand the coefficients of the metric. Using the Newtonian approximation to Kepler's third law, $r_e \dot{\phi}^2 \approx GM_\odot/r_e^2$, we find that the proper time interval measured by a stationary clock on Earth is related to eq.~\eqref{eq_Delta_t_final} via
\begin{equation} \label{eq_proper_time}
\Delta\tau \approx \left( 1 - \frac{3 GM_\odot}{2c^2 r_e} \right) \Delta t \, ,
\end{equation}
which gives a constant correction factor deviating from unity by\footnote{$r_e = 1 \, {\rm AU} = 1.49 \times 10^{11}$~m.} $\approx 10^{-8}$.

With eqs.~\eqref{eq_Delta_t_final} and~\eqref{eq_proper_time} the proper time elapsed between emission and arrival of the signal on Earth (round trip) is finally
\begin{eqnarray} \label{eq_Delta_tau_final}
\Delta \tau & = & \frac{2}{c} \left[ \sqrt{r_e^2 - r_0^2} + \sqrt{r_r^2 - r_0^2} \right] \left( 1 + \frac{\alpha^2}{2} \right) \nonumber \\
& - & \frac{3 GM_\odot}{c^3 r_e} \left[ \sqrt{r_e^2 - r_0^2} + \sqrt{r_r^2 - r_0^2} \right]    \nonumber \\
& + & \frac{4GM_\odot}{c^3} \left[  \sqrt{ \frac{r_e - r_0}{r_e + r_0} } + \sqrt{ \frac{r_r - r_0}{r_r + r_0} } \right] \left( 1 + \alpha^2 \right)  \\
& + & \frac{4GM_\odot}{c^3} \log\left[ \frac{ \left( r_e + \sqrt{r_e^2 - r_0^2} \right) \left( r_r + \sqrt{r_r^2 - r_0^2} \right)  }{r_0^2} \right] \, , \nonumber
\end{eqnarray}
where we neglected terms of order $\mathcal{O}(\alpha^2 GM_\odot/c^2 r_e)$, as well as post-post-Newtonian corrections. It is worth keeping in mind that the gravitational contributions are of order $GM_\odot/c^3 \approx 5 \, {\rm \mu s}$. Note that finite-mass corrections affect only the first and third terms, but crucially not the phenomenologically most relevant logarithmic term. 


Before closing this section, let us make a few general comments about our main result, eq.~\eqref{eq_Delta_tau_final}. The first term represents the Newtonian travel time in the absence of the Sun. This is effectively an Euclidean distance which cannot be measured directly due to the actual curvature of spacetime. The mass dependence of this term is expected, since the time it takes for a particle with normalized velocity $\beta = v/c$ to cover a distance $\ell$ is $(\ell/c)\beta^{-1}$. But in special relativity $E/m_\gamma c^2 = (1 - \beta^2 )^{-1/2} = \alpha^{-1}$, which can be rearranged as $\beta^{-1} = (1 - \alpha^2 )^{-1/2}$. Expanding this expression yields $\beta^{-1} \approx 1 + \alpha^2/2$, thus matching the mass-dependent factor in the first line of eq.~\eqref{eq_Delta_tau_final}. The other terms are general relativistic in nature and represent an extra travel time due to gravity. The gravitational time delay, the so-called Shapiro effect, is defined as the non-Newtonian contribution to the total round-trip time,
\begin{equation} \label{eq_tau_delay}
\Delta\tau_{\rm delay} = \Delta\tau - \Delta\tau(M_\odot = 0) \, .
\end{equation}
From now on we shall consider only the time delay as defined above.


\section{Bound on the photon mass via Cassini data}  \label{sec_exp_sens}
\indent

In Sec.~\ref{sec_time_delay} we obtained the correction to the round-trip travel time of a signal as measured by an observer on Earth, mediated by massive photons, to first post-Newtonian order, cf. eq.~\eqref{eq_Delta_tau_final}. In order to constrain a putative rest mass of the photon we must compare our theoretical model to observational data.

The best measurement of the Shapiro effect was performed using the Cassini spacecraft on its way to Saturn between 6 June and 7 July 2002, when the spacecraft was at heliocentric distances $r_r \approx 7.42$~AU\footnote{In this period Cassini's heliocentric distance varied very slightly, linearly going from $r_r = 7.38$~AU to $r_r = 7.46$~AU. Here we shall take the constant value $r_r = 7.42$~AU as a very good approximation. These data were obtained using the JPL Horizons on-line solar system data and ephemeris computation service: \url{https://ssd.jpl.nasa.gov/horizons/app.html}.}. During its superior conjunction on 21 June 2002 (13:14h UTC), the Cassimi spacecraft was at a geocentric distance of $8.4$~AU and the electromagnetic signal reached its closest approach to the Sun at $r_{\rm 0, min} = 1.6 \, R_\odot$. The radio link was setup so that essentially two carrier signals were sent to and from the spacecraft. The uplink had $f_u = 7.2$~GHz in the X-band and $f_u = 34$~GHz in the Ka-band, whereas the downlink had $f_d = 8.4$~GHz in the X-band and $f_d = 32$~GHz in the Ka-band. The possibility of using different frequencies allowed the almost complete elimination of coronal effects. Doppler-tracking of the signal was used to constrain $\gamma$ to unprecedented -- and so far unsurpassed -- accuracy: $|\gamma - 1| = 2.3 \times 10^{-5}$~\cite{Cassini_1, Cassini_2}.

Instead of trying to directly measure the time delays, the strategy was to observe the difference in frequency of the transmitted (T) and received (R) radar signals. More concretely, the main observable was the fractional frequency shift (FFS)~\cite{Cassini_1} (cf. eqs.~\eqref{eq_Delta_tau_final} and~\eqref{eq_tau_delay})
\begin{equation} \label{eq_y}
y = \frac{f_{\rm R}(\tau) - f_{\rm T}}{f_{\rm T}} = -\frac{d \left( \Delta\tau_{\rm delay} \right) }{d\tau} \, .
\end{equation}

\begin{figure}[t!]
\begin{minipage}[b]{1.0\linewidth}
\includegraphics[width=\textwidth]{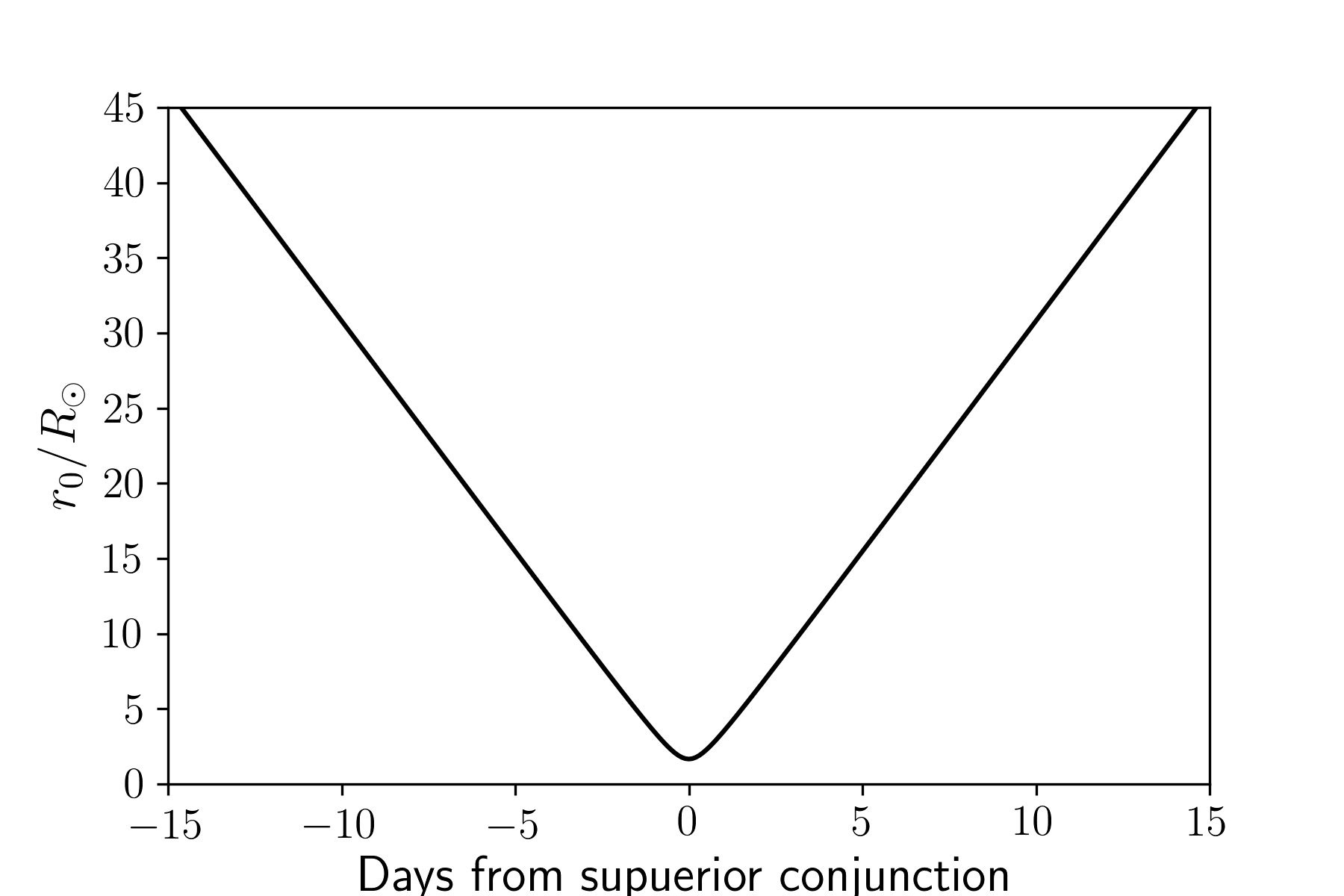}
\end{minipage} \hfill
\caption{Impact parameter as a function of time, cf. eq.~\eqref{eq_r0_tau} with $r_{0, {\rm min}} = 1.6 \, R_\odot$ and $v = 24.7$~km/s. The curve matches the one in Fig.~2 of Ref.~\cite{Cassini_imp_param}.}
\label{fig_r0}
\end{figure}

The data-taking period was roughly a month long and was centered around superior conjunction, so that, as time passes, the main time-dependent parameter is the point of closest approach of the radio signal to the Sun, $r_0 = r_0(\tau)$. Close to Cassini's superior conjunction, when the spacecraft, the Sun and Earth (in this order) were almost perfectly aligned, the variation of $r_0$ was not very different from Earth's orbital velocity of $30$~km/s. In fact, from Fig.~1 of Ref.~\cite{Cassini_2} we have 
\begin{equation} \label{eq_r0_tau}
r_0(\tau) = \sqrt{v^2 \tau^2 +r^2_{0, {\rm min}} } \, 
\end{equation}
with $r_{0, {\rm min}} = 1.6 \, R_\odot$, matching Fig.~2 of Ref.~\cite{Cassini_imp_param} for $v = 24.7$~km/s, cf. Fig.~\ref{fig_r0}. Now, $dr_0/d\tau = \mathcal{V}(r_0)$ can be written as
\begin{equation} \label{eq_vel}
\mathcal{V}(r_0) = v \sqrt{ 1 - \left( \frac{r_{0, {\rm min}}}{r_0} \right)^2 } \, ,
\end{equation}
so that
\begin{equation}  \label{eq_y_r0}
y = -\mathcal{V}(r_0)\frac{d\left(\Delta\tau_{\rm delay} \right)}{dr_0} \, .
\end{equation}
Note that $dr_0/d\tau < 0$ before superior conjunction and $dr_0/d\tau > 0$ after it, so the sign of $v$ must be changed accordingly.

\begin{figure}[t!]
\begin{minipage}[b]{1.0\linewidth}
\includegraphics[width=\textwidth]{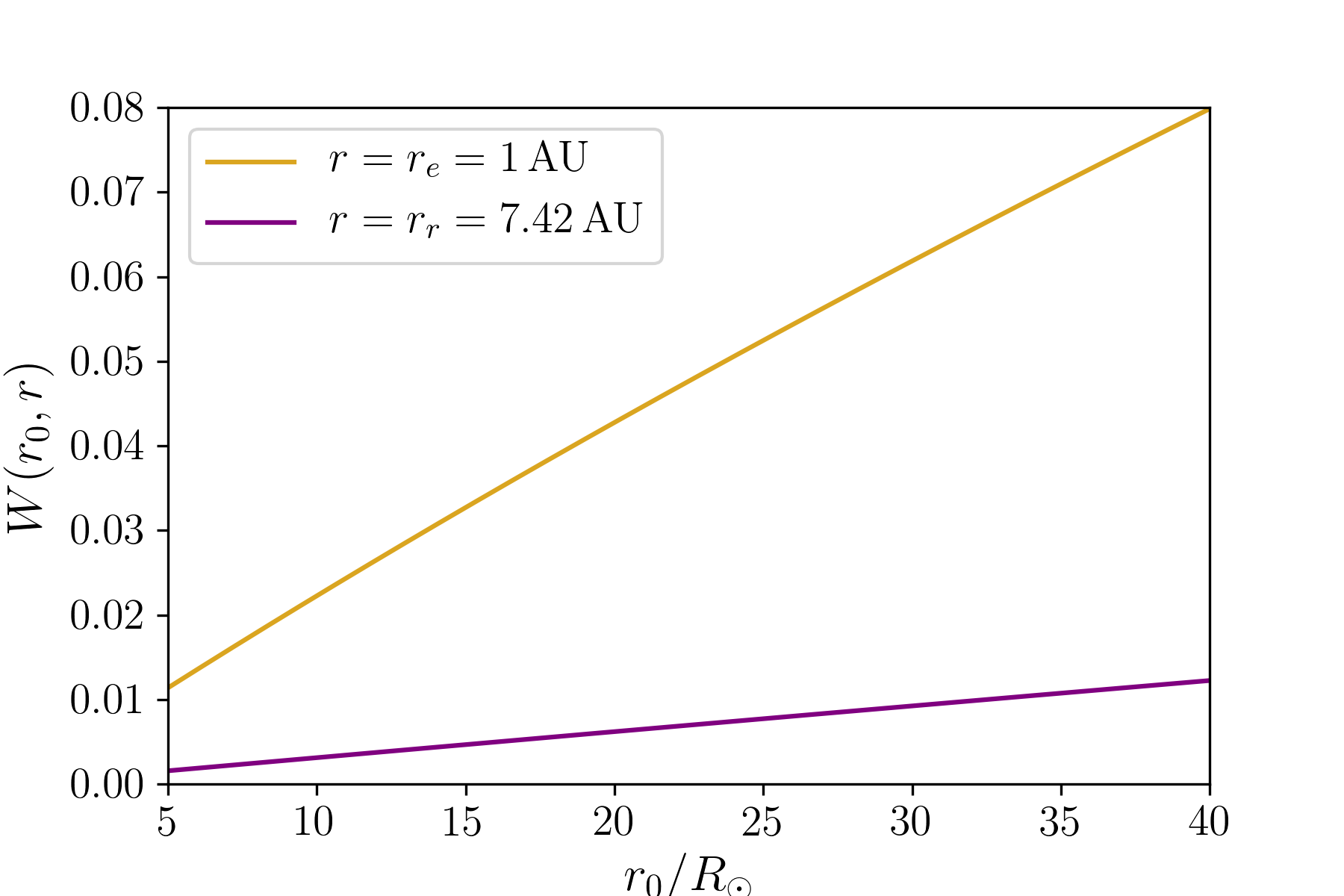}
\end{minipage} \hfill
\caption{The function $W(r_0, r)$, cf. eq.~\eqref{eq_W}, controls the magnitude of the mass-dependent correction to the FFS. In our current scenario, where $r_0 \ll r_{e,r}$, we have an approximately linear behavior given by $W(r_0, r) \approx 2.3 \times 10^{-3} (r_0/r)$ with $r_0$ expressed in units of $R_\odot$ and $r$ in AU. }
\label{fig_W}
\end{figure}

Our main result, eq.~\eqref{eq_Delta_tau_final}, can be split in two pieces: one with $\alpha$-independent terms equivalent to the standard predictions of general relativity and one $\alpha$-dependent. Inserting it into eq.~\eqref{eq_y_r0} the theoretical FFS becomes $y_{\rm th}(\alpha^2) = y_{0} + \alpha^2 y_\alpha$. The standard contribution is
\begin{equation}\label{eq_y_0} 
y_0 = \frac{8GM_{\odot} \mathcal{V}(r_0)}{c^{3}r_{0}} \left[ 1 + W(r_0, r_e) + W(r_0, r_r) \right]  \, ,
\end{equation}
where $8GM_{\odot}\mathcal{V}(r_0)/c^{3}R_\odot \lesssim 10^{-9}$. The leading term stems from the logarithm and the sub-dominant ones from the third line in eq.~\eqref{eq_Delta_tau_final}. These are controlled by the function
\begin{equation}  \label{eq_W}
W(r_0, r) = \frac{r_0 r}{2(r +r_{0})^{3/2}\sqrt{r -r_{0}}}    \, ,
\end{equation}
shown in Fig.~\ref{fig_W}. Since $W(r_0, r) \lesssim \mathcal{O}(10^{-2})$, we are allowed to keep only the leading contribution in $y_0$. Finally, the novel, mass-dependent correction to the FFS is
\begin{equation}\label{eq_y_alpha}
y_\alpha = \frac{8GM_{\odot} \mathcal{V}(r_0)}{c^{3}r_{0}} \left[  W(r_0, r_e) + W(r_0, r_r)  \right]  \, .
\end{equation}
To obtain these expressions we neglected quadratic terms in $r_0$ in eq.~\eqref{eq_Delta_tau_final}, since $r_0^2 \ll r_{e,r}^2$.

The analysis of the time-delay data performed in Ref.~\cite{Cassini_2} was based on the accurate characterization of Cassini's orbital dynamics. This was done using the Orbit Determination Program from NASA's Jet Propulsion Laboratory with which a fit of the data with up to 12 free parameters, including the initial state vector of the Cassini spacecraft, non-gravitational accelerations, thermo-optical properties of the antenna and Eddington's $\gamma$ parameter, was performed (see App.~\ref{app_D}). The orbital parameters were determined with uncertainties in position $\lesssim 50$~km for geocentric distances of $\approx 8.4$~AU. Furthermore, the uncertainty in the FFS, given by the Allan deviation, was found to be $\sigma_y \approx 2\times 10^{-14}$, practically constant and a factor $\approx 10^4$ smaller than the measured and predicted FFS~\cite{Cassini_1, Cassini_2, Cassini_imp_param, Cassini_error_budget}.

Testing new theories means that their free coefficients must be constrained using the available experimental data -- in our case by repeating the data analysis leading to the constraint on $\gamma$. This is particularly true for alternative metric theories of gravity (see {\it e.g.} Ref.~\cite{Hess}), in which the metric tensor and all its by-products, such as the geodesic equations and the resulting orbits, are modified. Massive electrodynamics, however, does not belong to this category: a massive photon, being exquisitely light, does not sufficiently deform spacetime -- it acts as a test particle moving in the gravitational field of the Sun and the Schwarzschild metric remains independent of $m_\gamma$. Consequently, the orbital dynamics of the Cassini spacecraft (weighing $\approx 5.2$~tons) is unaffected by a tiny, but finite photon mass and the available data may be used without a re-evaluation of the fits.

\begin{figure}[t!]
\begin{minipage}[b]{1.0\linewidth}
\includegraphics[width=\textwidth]{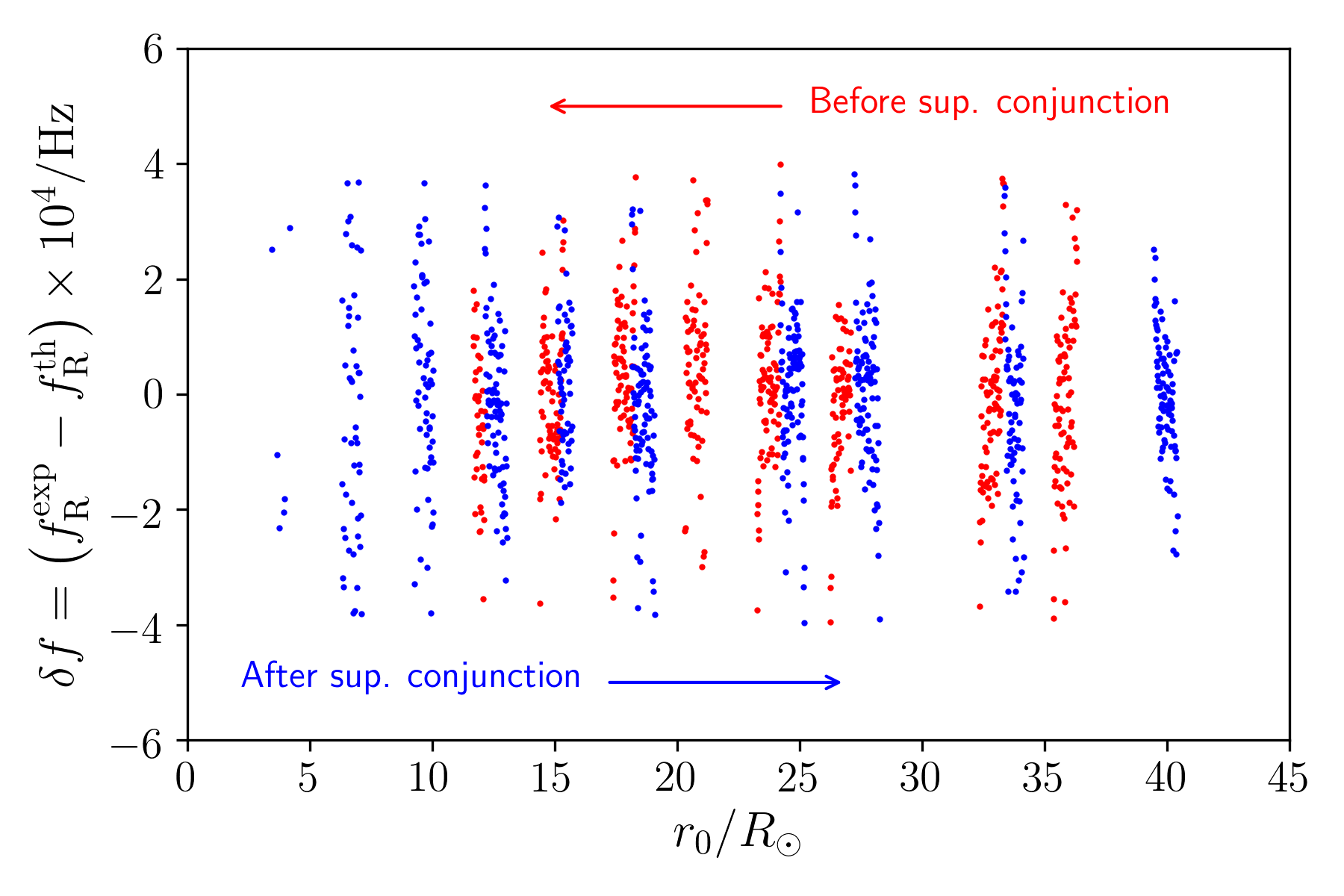}
\end{minipage} \hfill
\caption{Frequency residuals relative to a carrier frequency $f_{\rm T} =8.4$~GHz and respective impact parameters, courtesy of P. Tortora (U. Bologna). The arrows indicate the time sequence of the measurements from 6 June to 7 July 2002.}
\label{fig_df_r0}
\end{figure}

The data comprise $N = 1279$ Doppler (frequency) residuals in the form $\delta f = f_{\rm R}^{\rm obs} - f_{\rm R}^{\rm th}$, where the superscripts ``th" and ``obs" stand for theoretical and observed, respectively. These are normally distributed about zero with $|\delta f| \lesssim 5 \times 10^{-4}$~Hz relative to a carrier frequency $f_{\rm T} =8.4$~GHz. The data are shown in Fig.~\ref{fig_df_r0}. From eq.~\eqref{eq_y} we obtain the observed FFS as
\begin{equation} \label{eq_y_exp}
y_{\rm obs} = y_0 + \frac{\delta f}{f_{\rm T}} \, 
\end{equation}
that may now be used to evaluate the chi-square
\begin{equation} \label{eq_chi2}
\chi^2(\alpha^2) = \sum_{i=1}^{N} \frac{ \left[ y_{\rm obs}^i - y_{\rm th}^i(\alpha^2) \right]^2 }{ \sigma_y^2} \, 
\end{equation}
with $\sigma_y = 2 \times 10^{-14}$. The chi-square does not have a minimum for any finite value of $\alpha^2$: for any $\{ \alpha^2_1, \alpha^2_2 \}$ satisfying $\alpha^2_1 > \alpha^2_2$ we have $\chi^2(\alpha^2_1) > \chi^2(\alpha^2_2)$  with a limiting value $\chi_{\rm lim}^2/N \approx 0.687$. This indicates that the true minimum is at $\alpha^2 = 0$, that is, the data are compatible with massless electrodynamics, as expected.


We are thus only able to set upper bounds on $\alpha^2$. This may be done by searching for $\alpha^2 = \alpha^2_\star$ for which $\chi^2(\alpha^2_\star) - \chi_{\rm lim}^2 = 3.84$, corresponding to a $95\%$ confidence level~\cite{PDG}. We find $\alpha^2_\star = 1.9 \times 10^{-4}$ which translates to (cf. eq.~\eqref{eq_alpha})
\begin{equation} \label{eq_final_bound}
m_\gamma < 4.9 \times 10^{-7} \, {\rm eV/c^2} \, ,
\end{equation}
where we used $E = h f_{\rm T} = 0.035$~meV.

\section{Concluding remarks}  \label{sec_conclusions}
\indent

In this paper we analysed the gravitational time delay of massive photons. Using heliocentric isotropic coordinates we worked out the equations of motion and, following the actual geodesic path, we obtained the two-way (proper) time delay to first post-Newtonian order, cf. eq.~\eqref{eq_Delta_tau_final}. Using Doppler-tracking data from the 2002 Cassini experiment, we were able to constrain the photon mass at the level of $10^{-7} \, {\rm eV/c^2}$ at 95$\%$ confidence level.

As mentioned at the end of Sec.~\ref{sec_angle}, bounds of similar magnitude were reported in Refs.~\cite{Accioly1, Accioly2} by using measurements of the deflection angle of $2-43$~GHz radio waves. Though the frequencies used were roughly the same, the experimental uncertainties were larger, also reflecting on the inferior limits on $|\gamma - 1|$~\cite{Fomalont, Lebach}. Nevertheless, their best upper bound is at the same level as ours, cf. eq.~\eqref{eq_final_bound}. This is explained by looking at which terms of the standard massless expression are modified by the photon mass: in the case of the deflection angle the deviation from general relativity is $\sim \delta_{\rm E} \alpha^2$, cf. eq.~\eqref{eq_delta}. The situation with the Shapiro effect is unfortunately different, since the main phenomenological contribution, the logarithmic term in eq.~\eqref{eq_Delta_tau_final}, is not directly affected by finite-mass corrections. Therefore, the very precise measurements performed in the Cassini experiment were crucial to partly compensate for the reduced sensitivity to the photon mass.


Besides the obvious advantage of more precise measurements, one of the most important factors in improving the upper bound on the photon mass is the frequency of the signal. Assuming that $\delta f/f_{\rm T} \approx \sigma_y$ remain the same as in the Cassini experiment, from eq.~\eqref{eq_alpha} we see that our limits improve with decreasing frequency, an effect also reported elsewhere~\cite{Accioly1, Accioly2, LSW}. This is in fact expected, since smaller masses are related to larger de Broglie wavelengths, meaning that low-frequency phenomena are better probes to effects related to a very small, but nonetheless finite photon mass.

Looking towards the future, there are a few promising proposals, either already underway or planned, aiming at improving the current limits on various PPN parameters~\cite{Turyshev}. One example is the Mercury Orbiter Radioscience Experiment (MORE)~\cite{MORE, MORE_1} on board of the BepiColombo mission launched in 2018 and planned to start orbiting Mercury in 2026 ($r_r = 0.39$~AU). The objective is to better understand Mercury's internal structure, as well as to perform precision tests of general relativity. For the latter, a similar strategy to remove coronal effects as in the Cassini mission is followed and two carriers are used. With much improved ranging accuracy and an Allan deviation $\sigma_y \approx 10^{-16}$, it is projected that $| \gamma - 1 |\lesssim 2 \times 10^{-6}$ will be achieved. To roughly estimate the impact of these improvement on the upper limit of the photon mass, we may assume that the experimental data still match the predictions from general relativity (to the available precision) and $\delta f/f_{\rm T} \approx \sigma_y$ with $f_{\rm T} = 8.4$~GHz. This gives us $\alpha^2 y_\alpha \lesssim \sigma_y$, meaning that 
\begin{equation} \label{eq_sens}
m_\gamma \sim f_{\rm T} \sqrt{\sigma_y} \, ,
\end{equation}
and we may naively expect an improvement of roughly an order of magnitude over eq.~\eqref{eq_final_bound} just from the projected lower uncertainty.

Future space-based projects, such as the Gravitational Time-Delay Mission~\cite{Turyshev, GTDM, Dirkx},  will employ lasers, the current golden standard for ranging due to improved accuray and compactness of emitter and receiver systems~\cite{Dirkx}. This choice is sensible, as optical frequencies are less impacted than radio by coronal plasma effects. Unfortunately, the use of lasers is less than ideal in the search for better limits on the photon mass. From our rough estimate, cf. eq.~\eqref{eq_sens}, we see that only substantial improvements in the uncertainties -- or conditions which drastically enhance $y_\alpha$ (cf. eq.~\eqref{eq_y_alpha}) -- will be able to compensate for the $\sim 100$~THz frequencies of optical lasers. Therefore, in order to approach $m_\gamma \lesssim 10^{-14} \, {\rm eV/c^2}$, the realm of current bounds based on Earth-scale phenomena~\cite{PDG, Kroll, Malta}, at realistic error levels, technological limitations must be overcome to allow time-delay measurements at sub-MHz-frequencies in space.


\begin{acknowledgments}
We are grateful to P. Tortora for kindly providing the Cassini time-delay data and patiently clarifying how to use them, and J. Heeck for the support with the statistical analysis. We also thank the anonymous referee for very constructive criticism, M.O. Calv\~ao and L.T. Santana for discussions on the limit of geometric optics, as well as J.A. Helay\"el and G.P. de Brito for reading the manuscript. P.C.M. and C.A.D.Z. are indebted to Marina and Karoline Selbach, and Lalucha Parizek and Crispim Augusto, respectively, for insightful discussions. This work was financed in part by the Coordena\c{c}\~{a}o de Aperfei\c{c}oamento de Pessoal de N\'ivel Superior - Brasil (CAPES) - Finance Code 001. C.A.D.Z. is partially supported by Conselho Nacional de Desenvolvimento Cient\'ifico e Tecnol\'ogico (CNPq) under the grant no. 310703/2021-2, and by Funda\c{c}\~{a}o Carlos Chagas Filho de Amparo \`a Pesquisa do Estado do Rio de Janeiro (Faperj) under Grant no. E-26/201{.}447/2021 (Programa Jovem Cientista do Nosso Estado).
\end{acknowledgments}


\appendix

\section{The massless case: Eddington's $\gamma$ parameter and the PPN metric} \label{app_A}
\indent

In the main text we determined the gravitational time delay in the case of a massive photon following the geodesic path from emitter to receiver and back. This approach is often replaced by a simplified calculation using a straight line to approximate the true geodesic. Here we highlight the difference between the two, in particular for the standard case of a massless photon.

Given the many different alternative theories of gravitation, it is generally useful to devise a convenient parametrization to quantify possible deviations from general relativity -- the parametrized post-Newtonian (PPN) metric does just that~\cite{Poisson, Will_book, Will_exp}. The most relevant modification is that characterized by Eddington's $\gamma$ parameter, which appears in the metric coefficients as
\begin{equation} \label{eq_coeffs_AB_gamma}
g_{00} \approx 1 - \frac{2\mathcal{M}}{r} \quad {\rm and} \quad g_{ii} \approx 1 + \gamma\frac{2\mathcal{M}}{r}  \, ,
\end{equation}
where $\mathcal{M} = GM_\odot/c^2$ and the index $i$ runs over the spatial coordinates; in general relativity $\gamma = 1$.

Here we compare the calculations based on the straight-line approximation with the more rigorous derivation based on the geodesic. From now on we deal exclusively with massless photons and, for the sake of simplicity, we do not apply the correction factor to convert coordinate time into proper time, cf. eq.~\eqref{eq_proper_time}.

\subsection{Straight-line approximation} \label{app_A_1}
\indent

In the literature, most of the discussions of the Shapiro effect is done in terms of the usual Schhwarzschild coordinates using a straight line as an approximation for the curved path light takes. Let us constrain the motion to the $x-y$ plane, where emitter and receiver are at $(-x_e, r_0)$ and $(x_r, r_0)$, respectively, with $r_0$ being the point of closest approach of the light ray.

The photon moves in the $x$-direction and the line element is $ds^2 = 0 = g_{00} c^2 dt^2 - g_{xx} dx^2$, so that, to leading order in $\mathcal{M}/r$,
\begin{equation}
dt \approx \frac{1}{c} \left[ 1 + (1 + \gamma) \frac{\mathcal{M}}{r}  \right]
\end{equation}
with $r = \sqrt{x^2 + r_0^2}$. Integrating from the emitter to the receiver and multiplying by two we get 
\begin{equation}\label{eq_Delta_t_line}
\Delta t_{\rm line} = \frac{2(r_e + r_r)}{c} + \frac{4GM_\odot}{c^3} \frac{(1 + \gamma)}{2} \log\left( \frac{ 4r_e r_r  }{r_0^2} \right)
\end{equation}
in the case where $r_0 \ll x_e, x_r \approx r_e, r_r$. Note that the logarithmic term is the only relativistic correction to the Newtonian travel time.

\subsection{Geodesic solution} \label{app_A_2}
\indent

Let us now repeat the calculation leading to eq.~\eqref{eq_Delta_t_final}, but for a massless photon. We could start with the Lagrangean~\eqref{eq_lag_iso_coords}, but it is simpler to go another way. Since we are dealing with a massless photon, we cannot use the proper time as affine parameter. We may, however, use the coordinate time to parametrize the movement and use it as the line element such that
\begin{equation} \label{eq_new_constraint}
dt^2 = \frac{B(r)}{c^2 A(r)} \left( dr^2 + r^2 d\phi^2 \right)
\end{equation}
with $A(r) = g_{00}$ and $B(r) = g_{ii}$ given by eq.~\eqref{eq_coeffs_AB_gamma}. Restricting the motion to the equatorial plane $\theta = \pi/2$, we may define a new Lagrangean
\begin{equation}
L = \frac{B(r)}{c^2 A(r)} \left( \dot{r}^2 + r^2 \dot{\phi}^2 \right) \, .
\end{equation}
Again, $\phi$ is a cyclic coordinate and its associated Euler-Lagrange equation reads
\begin{equation} \label{eq_phi_dot_PPN_geo}
\dot{\phi} = \frac{K' \left(1 - \frac{2GM}{c^2 r} \right)}{r^2 \left(1+ 2\gamma\frac{GM}{c^2 r} \right)} \, .
\end{equation}
The integration constant is determined by plugging eq.~\eqref{eq_phi_dot_PPN_geo} into the metric~\eqref{eq_new_constraint} divided by $dt^2$ and evaluated at the perihelion, $r = r_0$, where $\dot{r} = 0$. We find
\begin{equation} \label{eq_KK}
K' = c r_0 \left( \frac{ 1 + 2\gamma\frac{GM}{c^2 r_0}  }{1 - 2\frac{GM}{c^2 r_0}   } \right)^{1/2}  \, .
\end{equation}

Inserting the results above in the constraint equation and isolating $\dot{r}$ we get (cf. eq.~\eqref{eq_coeffs_AB_gamma})
\begin{equation}
c \frac{dt}{dr} = \sqrt{\frac{B(r)}{A(r)}}  \Bigg\{ 1 - \left( \frac{r_0}{r} \right)^2 \frac{A(r)}{A(r_0)} \frac{B(r_0)}{B(r)}  \Bigg\}^{-1/2} \, ,
\end{equation}
which may be integrated similarly to eq.~\eqref{eq_dtdr_1} by changing variables to $q = r_0/r$ and expanding in $2\mathcal{M}/r_0 \ll 1$. The final result for the round trip with $r_0 \ll x_e, x_r \approx r_e, r_r$ is
\begin{eqnarray}\label{eq_Delta_t_geo}
\Delta t_{\rm geo} & = & \frac{2(r_e + r_r)}{c}  \nonumber \\
& + & \frac{4GM_\odot}{c^3} \frac{(1 + \gamma)}{2} \left[ 2 + \log\left( \frac{ 4r_e r_r  }{r_0^2} \right) \right] \, .
\end{eqnarray}
This result matches the one found by Weinberg~\cite{Weinberg}.

We may now contrast the PPN expressions for a massless photon obtained in the straight-line approximation, cf. eq.~\eqref{eq_Delta_t_line}, with that determined via the geodesic, cf. eq.~\eqref{eq_Delta_t_geo}. Clearly, the Newtonian travel time is the same in both, as it should; the post-Newtonian terms, however, are different. In particular, for $\gamma = 1$ we have a difference of $8GM_\odot/c^3 \approx 40 \, {\rm \mu s}$ between eqs.~\eqref{eq_Delta_t_line} and~\eqref{eq_Delta_t_geo} that is purely due to the paths followed.

\section{Geodesic equations in isotropic coordinates} \label{app_B}
\indent

We start with the interval line element~\eqref{eq_ds2} and define the Lagrangean $L = m_\gamma(ds/d\tau)^2$, whose action functional $S = \int L d\tau$ must be extremized. It may be explicitly written as
\begin{equation} \label{eq_lag_iso_coords}
\frac{L}{m_\gamma} = A(r) c^2 \dot{t}^2 - B(r) \left[ \dot{r}^2 + r^2 \left( \dot{\theta}^2 + \sin^2\theta \dot{\phi}^2 \right)  \right] \, ,
\end{equation}
where, as usual, the dot denotes differentiation with respect to proper time. Here the functions $A(r)$ and $B(r)$ are given by (cf. eq.~\eqref{eq_ds2})
\begin{equation} \label{eq_coeffs_AB}
A(r) =\left( \frac{1 - \frac{GM}{2c^2 r}}{1 + \frac{GM}{2c^2 r}}  \right)^2 \quad {\rm and} \quad B(r) = \left( 1 + \frac{GM}{2c^2 r} \right)^4  \, .
\end{equation}

The Euler-Lagrange equations for the variables $q = \{ t, r, \theta, \phi \}$ are
\begin{subequations}
\begin{eqnarray}
2A(r) c^2 \dot{t} & = & \epsilon \, , \label{eq_tdot} \\
2B(r)\ddot{r} + A'(r)c^2 \dot{t}^2 & = & B'(r) \left[ \dot{r}^2 + r^2 \left( \dot{\theta}^2 + \sin^2\theta \dot{\phi}^2 \right) \right]   \nonumber \\
& - & 2B(r)r \left( \dot{\theta}^2 + \sin^2\theta \dot{\phi}^2 \right)  \, , \label{eq_rdot} \\
B(r) r^2 \sin\theta\cos\theta\dot{\phi}^2 & = & \frac{d}{d\tau} \left[ B(r) r^2 \dot{\theta}  \right]  \, , \label{eq_thetadot} \\
-2B(r) r^2 \sin^2\theta \dot{\phi} & = & \Lambda \, . \label{eq_phidot}
\end{eqnarray}
\end{subequations}
The primes in eq.~\eqref{eq_rdot} indicate differentiation with respect to the argument of the function. Here $\epsilon$ and $\Lambda$ are constants stemming from the fact that $\{t, \phi \}$ are cyclical variables. These equations are not independent because of the constraint $g_{\mu\nu} \dot{x}^\mu \dot{x}^\nu = c^2$, which reads
\begin{equation} \label{eq_constraint}
A(r) c^2 \dot{t}^2 - B(r) \left[ \dot{r}^2 + r^2 \left( \dot{\theta}^2 + \sin^2\theta \dot{\phi}^2 \right)  \right] = c^2 \, .
\end{equation}

Let us start with eq.~\eqref{eq_thetadot} and consider movement in the equatorial plane $\theta = \pi/2$ with $\dot{\theta} = 0$. These choices satisfy this equation and imply that the trajectory will remain on this plane for all time. The equations above then reduce to 
\begin{subequations}
\begin{eqnarray}
2A(r) c^2 \dot{t} & = & \epsilon \, , \label{eq_tdot_2} \\
2B(r)\ddot{r} + A'(r)c^2 \dot{t}^2 & = & B'(r) \left( \dot{r}^2 + r^2 \dot{\phi}^2  \right) \nonumber \\
& - & 2B(r)r\dot{\phi}^2  \, , \label{eq_rdot_2} \\
A(r) c^2 \dot{t}^2 & = &  B(r) \left( \dot{r}^2 + r^2 \dot{\phi}^2  \right) + c^2 \, , \label{eq_constraint_2} \\
-2B(r) r^2\dot{\phi} & = & \Lambda \, . \label{eq_phidot_2}
\end{eqnarray}
\end{subequations}
If we move away from the Sun we have $A(r) \rightarrow 1$, so that eq.~\eqref{eq_tdot_2} becomes $2c^2 \dot{t} = \epsilon$, but now we are in flat Minkowski spacetime, where $\dot{t} = \gamma$, the Lorentz boost factor (not to be confused with Eddington's $\gamma$ parameter), thus $2\gamma c^2 = \epsilon$. We may then say that $\epsilon = 2E/m_\gamma$ with $E$ being the total energy of the massive photon. Moving on to eq.~\eqref{eq_phidot_2}, at large distances $B(r) \rightarrow 1$, so $-2r^2\dot{\phi} = \Lambda$; if $\Lambda = -2\ell/m_\gamma$ we recover the Newtonian expression for the (orbital) angular momentum, $\ell$.

With the constraint from eq.~\eqref{eq_constraint_2} we may ignore eq.~\eqref{eq_rdot_2} and, differentiating eqs.~\eqref{eq_tdot_2}, \eqref{eq_constraint_2} and~\eqref{eq_phidot_2} with respect to the proper time we finally obtain the geodesic equations for the coordinates $\{ t,r,\phi \}$:
\begin{subequations}
\begin{eqnarray}
\ddot{t} & = & -\frac{2GM}{c^2 r^2} \frac{\dot{r}\dot{t}}{ \left( 1 - \frac{GM}{2c^2 r} \right) \left( 1 + \frac{GM}{2c^2 r} \right) } \, , \label{eq_geo_t} \\
\ddot{\phi} & = & -\frac{2}{r} \frac{\left( 1 - \frac{GM}{2c^2 r} \right)}{  \left( 1 + \frac{GM}{2c^2 r} \right) } \dot{r}\dot{\phi} \, , \label{eq_geo_phi} \\
\ddot{r} & = & -\frac{GM}{c^2 r^2} \frac{\left( 1 - \frac{GM}{2c^2 r} \right)}{  \left( 1 + \frac{GM}{2c^2 r} \right)^7 } \dot{t}^2 + 
\frac{\left( 1 - \frac{GM}{2c^2 r} \right)}{  \left( 1 + \frac{GM}{2c^2 r} \right) } r \dot{\phi}^2 \nonumber \\
& + &  \frac{GM}{c^2 r^2} \frac{1}{  \left( 1 + \frac{GM}{2c^2 r} \right) } \dot{r}^2 \, . \label{eq_geo_r} 
\end{eqnarray}
\end{subequations}
%

\section{Deflection angle in the straight-line approximation} \label{app_C}
\indent

In Sec.~\ref{sec_angle} we based our calculation on the geodesics followed by a massive, ultra-relativistic particle passing near a massive body. Here we repeat the calculations under the assumption that the geodesics may be approximated by straight lines. For simplicity, we take the trajectory to lie on the $x-y$ plane. In this case we have $M_\odot$ at $y = 0$ and the trajectory is taken to be parallel to the $y$-axis with $dx^\beta \approx (c/v,0,1,0)dy$, where $v$ is the velocity of the massive particle with 4-momentum $p^\alpha = (E/c, 0, p^y, 0)$~\cite{Nieves}.

Following Refs.~\cite{Ohanian, Nieves} we assume that the massive photon is moving along the positive $y$-direction with impact parameter $b$ along the $x$-axis. To first order in $GM_\odot/b c^2$, the total change in its 4-momentum is 
\begin{equation} \label{eq_delta_p}
\Delta p_\mu = \frac{1}{2} p^\alpha \int_{-\infty}^{\infty} h_{\alpha\beta,\mu} \, dx^\beta \, ,
\end{equation}
where $h_{\alpha\beta}$ is the deviation of the metric from flat Minkowski spacetime. Assuming a weak field, we have $h_{00} = h_{ii} \approx  -2GM_\odot/c^2 r$, cf. eq.~\eqref{eq_ds2}, and the change in 3-momentum in the $x$-direction is 
\begin{equation} \label{eq_px}
\Delta p_x = \frac{2GM_\odot}{c^2 b} \left( \frac{E}{v} + p^y \right) \, .
\end{equation}

Energy and momentum are conserved and we are evaluating the deflection in asymptotically flat regions, where the dispersion relation $E^2 = m_\gamma^2 c^4 + (cp^y)^2$ holds. Furthermore, energy and velocity are related via $E = m_\gamma c^2 / \sqrt{1 - v^2/c^2}$. Since $m_\gamma c^2 \ll E$, we have 
\begin{equation} \label{eq_E_pz_plus}
\frac{E}{v} + p^y \approx \frac{2E}{c} \left( 1 + \frac{\alpha^2}{2} \right) \, ,
\end{equation}
where $\alpha = m_\gamma c^2/E$. Noting that the deflection angle is given by $\delta = |\Delta p_x/p^y | \approx |\Delta p_x/(E/c)|$, we finally obtain
\begin{equation} \label{eq_def_angle_1}
\delta \approx \frac{4GM_\odot}{c^2 b} \left( 1 + \frac{\alpha^2}{2} \right) \, ,
\end{equation}
which matches Refs.~\cite{Accioly1, Accioly2}, as well as our more rigorous result~\eqref{eq_delta} derived using the full geodesic equations, provided we have $b \approx r_0$.

\section{Impact of a photon mass on the orbit determination of the Cassini spacecraft and parameter fit} \label{app_D}
\indent

The mass of the photon, $m_\gamma$, is small enough to make the photon a test particle in the solar gravitational background: the spacetime metric is independent of $m_\gamma$. The consequence is that the geodesics remain independent of $m_\gamma$. Therefore, on the theoretical side, the determination of the orbital parameters by means of solving the equations of motion is unaffected by $m_\gamma$. We are nonetheless dealing with a spacecraft moving at interplanetary distances whose trajectory must be precisely controlled to keep it on track.

The communication between spacecraft and ground station is performed using electromagnetic signals. It is thus plausible that a finite photon mass will have some impact on the measurement and subsequent reconstruction of the orbital parameters. As will be shown, the range and range rate are influenced by a finite $m_\gamma$, since these depend on the two-way light time. Let us first briefly address the orbit determination procedure.

Orbit determination is an iterative process starting with the nominal orbit involving the simultaneous integration of the equations of motion of all relevant participants (including all known influences)~\cite{Moyer, Thornton}. The determination of the trajectory is performed by comparing observed quantities, obtained via tracking (range and range rate) to calculated ones. Using suitable fitting procedures the optimal model parameter set is found and the residuals are minimized. The parameters estimated in this way are fed back into the model and the process is repeated until adequate convergence is achieved.

The model parameters (a total of 12) that are estimated in the fit are~\cite{Cassini_1, Cassini_2}: the six components of the spacecraft state vector at a reference epoch, the three components of the induced acceleration from the radioisotope thermal generators, the specular and diffuse reflectivity of the spacecraft high gain antenna, and the relativistic PPN parameter $\gamma$. If we were to repeat the orbital determination procedure, we would fix $\gamma = 1$ and include the photon mass as a free parameter.

Range and range rate (Doppler) measurements inform about distance and velocity relative to the tracking station on Earth. The spacecraft range is measured using the transit time of a ranging signal, $\Delta\tau$, which is sent from a ground station, electronically processed on-board and re-emitted to Earth. The range, $\rho$, is given by $\rho = c \Delta\tau$, whereas the range rate, $\dot{\rho}$, is determined via Doppler tracking: $f_{\rm R} = f_{\rm T} \left( 1 \pm \dot{\rho}/c \right)$, where $f_{\rm T}$ and $f_{\rm R}$ are the transmitted and received frequencies, respectively (the plus or minus sign depends on the sign of the relative range rate). Due to the dependence on the travel time (one way or two way), the modelled distances and velocities entering the orbital determination fit are potentially affected by a finite photon mass. This issue will be discussed below.

Frequencies are measured by comparison with hydrogen masers providing an extremely stable time standard. Even if a finite photon mass would have sensible effects on the hydrogen spectrum, the sent and received frequencies are both measured against the same reference frequency and the Doppler shifts are not large enough to produce meaningful effects. Therefore, the photon mass does not play a role in frequency comparison.

The re-transmission of the radio signal is performed as follows. The signal received at the spacecraft is already Doppler shifted (one-way Doppler shift). The transponder on-board, the Ka band translator, KaT, is responsible for collecting the incoming radio signal, measure its frequency, lock it in and then re-emit a signal with this one-way Doppler-shifted frequency back to Earth. The signal received at the ground station is therefore the two-way Doppler shifted signal, which contains the expected gravitationally delayed frequency, but also several other effects that must be removed, such as non-gravitational accelerations, coronal plasma effects and tropospheric effects. Despite the fact that some of these are measured using electromagnetic radiation (atmospheric measurements use radar), the very short times and distances involved lead us to not expect any sensitivity to a finite photon mass. Therefore, also in the re-transmission process the photon mass will have no impact.

Regarding non-gravitational accelerations, the only forces with potential impact on the experiment were the solar radiation pressure and the thermal emission from the radioisotope thermal generators~\cite{Cassini_error_budget}. The radiation pressure, which would have been affected by a photon mass, induces a non-gravitational acceleration a hundred times smaller than that due to thermal emissions, which were themselves found to be very small. These two effects are overall negligible and any correction from a finite photon mass would not change that.


We have assumed that a finite mass would have no meaningful impact in the orbital parameters, thereby justifying the usage of the Cassini data as presented. This approach would be strictly rigorous if the effects of a finite photon mass would be smaller than the errors in the determination of the orbital parameters, such as heliocentric or geocentric distance of the Cassini spacecraft.

The current best limit is $m_\gamma < 10^{-18} \, {\rm eV/c^2}$~\cite{PDG, Ryutov}, which gives $\alpha^2 < 10^{-28}$, cf. eq.~\eqref{eq_alpha} with $f_{\rm T} = 8.4$~GHz. From $v = c \sqrt{1 - \alpha^2}$ we may write $v = c - \Delta c$ with $\Delta c = c\alpha^2/2$ and find	
\begin{equation} \label{eq_delta_c}
\frac{c - v}{c} =\frac{\Delta c}{c} \approx 10^{-28} \, .
\end{equation}
Measuring this fractional difference is impossible with current or near future capabilities.

The orbital distances have been determined with unprecedented accuracy. The spacecraft position was determined with errors $\Delta L \approx 50$~km at geocentric distances of $L \approx 8$~AU~\cite{MORE},. We thus have
\begin{equation} \label{eq_delta_L}
\frac{L - (L \pm \Delta L)}{L} = \pm \frac{\Delta L}{L} \approx \pm 10^{-8} \, .
\end{equation}
Despite being an impressive feat, this is several orders of magnitude larger than the order of magnitude obtained in eq.~\eqref{eq_delta_c}.

How large would $m_\gamma$ have to be to produce a similar relative variation? Assuming that range measurements (or determinations via fits) would be affected by a finite photon mass, we may approximately write $\Delta L /L \approx \alpha^2$. Therefore, if the uncertainty in the position determination remains roughly the same, we must have $\alpha^2 < 10^{-8}$, which leads to $m_\gamma < 10^{-9} \, {\rm eV/c^2}$, roughly hundred times smaller than our upper bound. One could thus say that, had we used the full orbital determination program including all the dynamical variables used in the Cassini study, but now with the extra free parameter being $m_\gamma$ instead of $\gamma$, the fit could have (optimistically) returned an upper bound at this ballpark. Note that this takes only the position uncertainty into account, but there might be other larger sources of uncertainty, which would cause this limit to degrade further.


In conclusion, our bound is a conservative estimate. A more thorough -- but technically out of scope -- treatment could lead to a better limit by a factor $\sim 100$, probably less. As shown above, the heliocentric distances and other orbital parameters extracted from NASA's Horizon calculator and from the Cassini solar conjunction experiment are not expected to significantly change (within the stated relative uncertainties in range) if $m_\gamma < 10^{-9} \, {\rm eV/c^2}$. Our approach does not consider these tighter relative uncertainties, thus leading to the upper bound $m_\gamma < 4.9 \times 10^{-7} \, {\rm eV/c^2}$, representing a conservative limit that is compatible with the larger relative uncertainties in the FFS after the nominal fit was performed. This nonetheless encompasses the estimate given above. Therefore, we believe that there is a chance that our bound could improve by at most a couple of orders of magnitude if a re-fit would be feasible.



\end{document}